# Physics of multi-GEM structures

A. Buzulutskov *

*Budker Institute of Nuclear Physics, 630090 Novosibirsk, Russia*

**Abstract**

We show that physics of multi-GEM structures is rather complex, regarding the number of phenomena affecting detector performance. The high-pressure operation in noble gases and the ion feedback are considered in more detail. It is proposed that the dominant avalanche mechanism in He and Ne, at high pressures, is the associative ionization. Ion feedback effects related to the dependence on gas, pressure and gain and to possible avalanche extension outside the GEM holes are discussed.

*PACs:* 29.40C; 07.85.F.

## 1. Introduction

In recent years a considerable progress has been made in Gas Electron Multiplier (GEM) [1] detectors, motivated by the growing interest in their potential applications. The advanced features of GEM-based detectors are high counting rate, excellent spatial resolution, good imaging capability, operation in magnetic field, large sensitive area, flexible geometry and low cost [2].

A unique property of GEMs, as compared to other micro-pattern detectors, is their capability to operate in cascade, i.e. in a multi-GEM structure. The advantages of multi-GEM detectors comprise: effective photon [3,4] and ion feedback suppression [5-7]; high gains, reaching $10^7$ [3,4,8]; high gain (above $10^5$) operation in pure noble gases [3] including that at high pressures [9,10]; discharge prevention in intense particle beams [11]; optical readout directly on CCD-based sensors [12]; good time resolution, down to 1 ns for single electron counting mode [4]; operation in a sealed mode [6]; reduced ageing rate [6,13], operation with CsI-coated GEMs [14], etc.

The remarkable properties of multi-GEM structures are very attractive for the numerous applications: tracking detectors for high energy physics experiments [15], GEM-based gas photomultipliers [6], non-ageing detectors operated with pure noble gases [3,9,10], endcap detectors for time-projection chambers (TPC)

---

* Corresponding author. Fax: 7-3832-342163. Phone: 7-3832-394833.
Email: buzulu@inp.nsk.su

[16], cryogenic detectors for solar neutrino and dark matter search [17,18], windowless RICH detectors [19], imaging detectors with electrical [20] and optical [12] readout, fast detectors for triggering [21], neutron detectors [22], etc.

Such a broad variety of applications is based on various physical effects taking place in multi-GEM structures. In this paper we describe some of them and in particular the high-pressure operation in noble gases, which has a fundamental meaning from the point of view of electron avalanching in dense noble gases. The effects of ion feedback and avalanche extension from the GEM hole are also discussed.

## 2. Physics of multi-GEM structures

A typical multi-GEM detector consists of a cathode, a cascade of 3-4 GEM elements and a printed-circuit-board (PCB) anode mounted inside the gas volume. The GEM is made of a 50 μm thick Kapton foil, copper clad from both sides and perforated by a matrix of micro-holes. The hole pitch is 140 μm; the hole diameter is typically 70÷80 μm on the copper side and 40÷60 μm on the Kapton side. The drift gap (between the cathode and the 1$^{st}$ GEM), transfer gaps (between the GEMs) and induction gap (between the last GEM and the anode) are typically 3, 1÷2 and 1÷2 mm, correspondingly. The GEM electrodes are biased through a resistive high-voltage divider.

The physics of multi-GEM structures is rather complex, regarding the number of phenomena affecting the detector performance. A particular large number of effects is observed in multi-GEM photomultipliers operated with pure noble gases [3,23]. Fig.1 schematically illustrates some processes that might occur in such a detector :

- photoelectron backscattering to the photocathode, resulting in the reduction of the photocathode quantum efficiency in gas compared to that in vacuum [3,4];
- photon feedback to the photocathode induced by avalanche emission and secondary scintillations, resulting in the after-pulses [3,4];
- ion feedback from a GEM to the preceding GEMs and to the photocathode, resulting in the after-pulses and gain limitations due to ion-induced electron emission [3,23];
- ion feedback to the bottom GEM electrode, resulting in charging-up of kapton surfaces at high anode currents and leading to gain instabilities [3,23];
- avalanche confinement in the GEM holes, resulting in the avalanche saturation at relatively moderate gains [3,23];
- avalanche extension outside the GEM holes at high gains, resulting in the limitation on the minimum inter-GEM distance [3,23].

Other interesting physical effects relevant to the multi-GEM structure performance are the discharge propagation [11], GEM electron transparency [5,24], high-pressure operation in traditional [25] and pure noble gas mixtures [9,10], low-pressure operation [26], ageing [6,13], electric field pattern computation [5], etc. Some of these topics are considered in the following sections.

## 3. High-pressure operation in noble gases

High-pressure operation of GEM detectors in pure noble gases is relevant in the field of double-phase cryogenic particle detectors [27] for solar neutrino and dark matter search [17,18]. In such detectors the primary ionization produced in liquid He, Ne or Xe is extracted into the gas phase where it is detected using the GEM multiplier. Since the gas density is a reciprocal function of temperature, at a given pressure, the gas detector operation at cryogenic temperatures at atmospheric pressure is equivalent to that at high pressures at room temperature. In addition, the high-pressure operation in pure He is attractive for He$^3$-based neutron detectors [22], X-ray imaging detectors with solid convertors [28], high-pressure helium TPC for solar neutrino detection [29] and colliding beam physics.

The experimental results on the triple-GEM detector operation in compressed noble gases were obtained in [9,10]: Fig.2 shows the maximum detector gain as a function of pressure and Fig.3 shows the pressure dependence of the maximum operation voltage that can be applied across each GEM ("discharge voltage"). In Fig.3 the data obtained in Ar/$CO_2$ and $CH_4$ with the single-GEM detector [25], are presented for comparison.

In contrast to molecular gases, the discharge voltages in practically all noble gases stop increasing when exceeding 3 atm. We relate this to a specific nature of the discharge mechanism in noble gases: the discharges are presumably generated by ion-induced electron emission due to the ion feedback from the last to preceding GEMs [3]. This is because the ion-induced electron emission is considerably enhanced in noble gases compared to other gases [33]. On the other hand, the operation voltages in Ar, Kr and Xe increase with pressure, reflecting the $E/p$ dependence of the ionization coefficient. Therefore the maximum gain in these gases drops rapidly for pressures exceeding 3 atm.

The operation in light noble gases is different from that of heavy noble gases: their operation voltages grow very slowly or even decrease with pressure [9,10]. This may explain the large gain values obtained in He and Ne at high pressures, reaching $10^5$, despite of the fact that their operation voltages are by a factor of 2-3 lower than those in Ar, Kr and Xe.

In order to understand the avalanche mechanism in noble gases at high pressures, we estimated the Townsend ionization coefficients, $\alpha_H$, using the experimental data and compared them to those available in the literature. Following ref. [30], a parallel-plate approximation to the avalanche development inside the GEM hole was applied. In this simplified approach the avalanche is considered to develop in the uniform electric field inside the hole, $E_{GEM}$, over the distance $d=50$ μm equal to the GEM inter-electrode distance. The field $E_{GEM}$ is taken to be equal to the computed field in the center of the hole: $E_{GEM}$ =63 kV/cm at the GEM voltage $\Delta V_{GEM}$=500 V [5]. The gain of a single GEM, $M$, is determined from the total triple-GEM gain, $G$, measured in experiment, using the formula $G = (M\varepsilon)^3$, where the charge transfer efficiency from the GEM output to the following elements, $\varepsilon$, was taken to be 1/3 [5]. Then we have

$$M = \exp(\alpha_H d) \quad \Rightarrow \quad \alpha_H / p = \ln M / (pd)$$

Fig.4 shows the comparison of the reduced ionization coefficients estimated in this way, $\alpha_H / p$, to those compiled from the literature [31-35], $\alpha_L / p$, in He, Ne and Ar. Since there is a discrepancy between various presentations of ionization coefficients for Ne and He, in particular at low $E/p$ values, we chose the presentations that gave the largest $\alpha_L / p$ values [31,32]. It should be emphasised that these ionization coefficients were measured at low pressures, namely below 80 Torr in He [32] and below 150 Torr in Ne [31]. Therefore in fact Fig.4 provides the comparison between the data obtained at high (1-15 atm) and low (below 0.2 atm) pressures.

One can see that in He and Ne the ionization coefficients are considerably larger at high pressures than at low pressures. Moreover, the data obtained at different pressures strongly violate $E/p$ scaling behaviour. That means that at high pressures the reduced ionization coefficient becomes dependent not only on the reduced electric field, as it is in the electron impact ionization mechanism, but on the pressure as well. At the same time there is a relatively good agreement between high and low pressure data in Ar, Kr and Xe (Kr and Xe data are not shown). These results may indicate that a new avalanche mechanism arises at high pressures in light noble gases, other than the electron impact ionization.

Most probably this mechanism is the associative ionization [34-36]. In the associative ionization, the electron is produced in atomic collisions due to the association of an atom with an excited atom into a molecular ion: $He + He^* \to He_2^+ + e$. In He, this reaction goes via short-lived excited atomic states, the lowest energy levels of which are 3P and 3D. The reaction cross-section is rather large, of about $10^{-15}$–$10^{-16}$ cm$^2$. Also, the energy threshold for the molecular ion appearance is lower than that of the atomic ion, by about 1 eV, in all noble gases.

The hypothesis that the associative ionization is responsible for the discrepancy in ionization coefficients observed in He at pressures below 1 atm, as well as for the $E/p$ scaling violation in $\alpha/p$, was introduced in [35]. It is interesting that according to theoretical calculations [37] the avalanche development in liquid He, at low electric fields, would also be defined by the associative ionization. In addition, we show below that the contribution of the associative ionization might provide the required pressure dependence.

**4. A model for avalanche development in noble gases**

Let us consider a simplified model for the avalanche development including the following basic processes [36]: impact ionization, atomic excitation, associative ionization and de-excitation due to photon emission, with the corresponding reaction rate and time constants $k_i$, $k_e$, $k_r$ and $\tau$:

$e + A \xrightarrow{k_i} A^+ + 2e$, impact ionization;

$e + A \xrightarrow{k_e} A^* + e$, excitation;

$A^* + A \xrightarrow{k_r} A_2^+ + e$, associative ionization;

$A^* \xrightarrow{1/\tau} A + \hbar\omega$, de-excitation.

This model is described by the following system of equations [36]:

$$\frac{dN_1}{dt} = N_e N_a k_i \quad ; \quad \frac{dN_2}{dt} = N^* N_a k_r \quad ;$$

$$\frac{dN^*}{dt} = N_e N_a k_e - N^*(\frac{1}{\tau} + N_a k_r) \quad ;$$

$$N_e = N_1 + N_2 = \exp(\nu t) \quad ,$$

where $N_e$, $N_a$, $N_1$, $N_2$, $N^*$ are the densities of electrons, atoms, atomic ions, molecular ions and excited atoms, respectively; $\nu$ is the total frequency of ionization.

The production of ions in the impact ionization increases in proportion to the pressure. An additional pressure dependence, which we would like to obtain, is provided only if the de-excitation time constant is smaller than the collision time for the associative ionization. Indeed, if it is not the case, the excited atom would produce the molecular ion anyway, independently of the collision rate and thus of the pressure, like it would be for long-lived metastable atoms (the Penning effect).

Thus, solving the system of equations when the de-excitation frequency is higher than the ionization frequency, i.e. when $1/\tau \gg \nu$, we obtain the ratio of ionization coefficients for the impact ($\alpha_i$) and associative ($\alpha_r$) ionizations:

$$\frac{\alpha_r}{\alpha_i} = \frac{dN_2/dt}{dN_1/dt} \approx \frac{k_e}{k_i} k_r \tau N_a \approx const \cdot p \quad .$$

Here we supposed that in the first approximation the ratio of excitation to impact ionization rate constants, $k_e/k_i$, is independent of the electric field. One can see that due to the contribution of the associative ionization, the total ionization coefficient, $\alpha_t = \alpha_i + \alpha_r$, becomes dependent on both the electric field and the pressure:

$$\frac{\alpha_t}{p}(\tfrac{E}{p}, p) \approx [1 + const \cdot p]\frac{\alpha_i}{p}(\tfrac{E}{p}) \quad .$$

To check this relation, the quantity $\alpha_H/p/(1+Cp)$ is fitted to $\alpha_L/p$, where the fitting parameter $C$ describes the contribution of the associative ionization at high pressures: see Fig.5.

One can see that the agreement between high and low pressure data in He and Ne is achieved only if to take into account the associative ionization contribution. This contribution is about 2/3 in He and ½ in Ne at atmospheric pressure; it fully dominates at higher pressures, but becomes negligible at pressures lower than 0.2 atm, i.e. just in the region where the ionization coefficients of noble gases, presented in the literature, were measured. In Ar, Kr and Xe the associative ionization is negligible at all pressures.

The associative ionization is a fast process with a time constant of the order of a nanosecond at atmospheric pressure [35]. It goes via basic and short-lived excited states of the native atoms. Therefore other possible processes like those going via collisions with atoms of impurities would have a minor contribution due to the lower atomic concentration. That means that the high-pressure operation of avalanche detectors in light noble gases should not be very sensitive to impurities, in contrast to low-pressure operation and in accordance with recent observations [10,38].

There might also exist the ionization processes going via long-lived meta-stable atomic states, such as the Penning ionization of impurities [34-36]. Just due to the same reason, it would also have a minor effect at high pressures, against the associative ionization background. In addition, its pressure dependence would be similar to that of the impact ionization, as was pointed out above and observed in experiment [31].

### 5. Ion feedback

Ion feedback studies in multi-GEM structures are important for understanding discharge mechanisms in noble gases, prevention of field distortion in TPCs [16] and reduction of ion-induced electron emission from the photocathode in gas photomultipliers [6]. The results of the detailed study of ion feedback are presented elsewhere [7]. Since the mechanism of ion feedback suppression in multi-GEM structures is not fully understood, we consider here only those effects that might have some apparent explanations: the influence of the gas and pressure and the possible connection to the effect of the avalanche extension from the GEM holes.

Fig.6 shows the ion feedback fraction (the ratio of the cathode-to-anode currents) as a function of the gain of a triple-GEM detector, in He at 1, 5 and 10 atm and in Ar/CF$_4$ at 1 atm [7]. Most of the data were obtained when the voltage across the drift gap was equal to that applied across each GEM, i.e. when the drift field was proportional to electric field inside the GEM hole. One data set, in Ar/CF$_4$, is obtained at a constant drift field: $E_D$=0.5 kV/cm.

The results might seem to be surprising. Indeed, the difference in operation voltages between He and Ar/CF$_4$ is of a factor of 2-2.5 and the difference in $E/p$ values reaches a factor of 10 (see Figs.2,3). Despite of this, the ion feedback is practically independent of the gas, pressures and $E/p$, for a given gain (fig.6). The conclusion might be that the electron and ion diffusion, which is generally a function of the pressure, gas and reduced electric field, does not affect the ion feedback. That means that the ion feedback is determined by something else, probably by spatial distributions of the avalanche and the electric field in the GEM hole.

Another interesting feature is that the ion feedback fraction $F$ is with good accuracy an inverse power function of the gain $G$, in a wide gain range (Fig.6):

$F = aG^{-b}$ ; $a, b > 0$ ;

where parameters $a$ and $b$ are independent of the gas and the pressure. The future theory of ion feedback should be able to explain this dependence.

The intriguing observation in Ar/CF$_4$ is that starting from a certain critical gain, of about $5 \times 10^4$, the ion feedback suppression is considerably enhanced. We suppose that this is connected to the effect of the avalanche extension outside the GEM holes [3], which has a threshold dependence on gain [23]. Indeed, if it would be the case, the positive ions produced outside the hole would have more chances to be terminated at the bottom GEM electrode rather than to go through the hole. It is interesting that the critical gain, at which the supposed avalanche extension takes place here, is of the same order as that estimated earlier in Ar from the pulse-shape analysis [23]: $4 \times 10^4$. This speaks in favour of the avalanche extension hypothesis.

### 6. Conclusions

The physics of multi-GEM structures is rather complex; the performance of such structures is characterized by a large number of physical effects. A few of them, namely the high-pressure operation in noble gases, ion feedback and avalanche extension, were considered in this work.

In order to describe the unusual pressure dependence of avalanche characteristics in light noble gases, He and Ne, it is proposed that the avalanche mechanism in these gases, at pressures exceeding 1 atm, is due to the associative ionization. On the other hand, the data in Ar, Kr and Xe are described well by the electron impact ionization mechanism. The question why such a difference exists between light and noble gases should be addressed to more elaborated theory. The practical consequence of the proposed avalanche mechanism is that the operation voltages in multi-GEM detectors are much lower than expected, allowing for high gain operation in dense He and Ne. This property is very attractive for numerous applications in high energy physics, astroparticle detection, nuclear physics and medical imaging.

The detailed mechanism of ion feedback suppression in multi-GEM structures is not yet well understood. In particular, the inverse power dependence of ion feedback fraction on gain, observed in experiment, should be explained in the future. The independence of the pressure and the gas may indicate that the ion feedback is governed mainly by geometrical factors inside the GEM hole, such as the hole diameter and shape, the electric field pattern and the avalanche shape. In particular, the change of the avalanche shape due to its possible extension outside the GEM hole might explain the enhanced suppression of ion feedback observed at high gains.

Further studies of these and other effects in multi-GEM structures, e.g. operation in noble gases at cryogenic temperatures, in a gas and liquid phase, light emission in an avalanche, double-phase detectors, etc., are in course.

### Acknowledgements

The author is indebted to Drs. A. Bondar and L. Shekhtman of the Budker Institute for useful discussions.

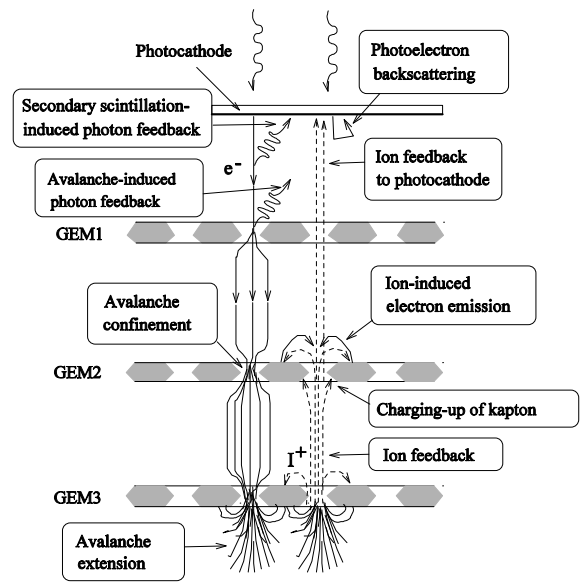

Fig.1 Physical effects taking place in multi-GEM photomultipliers operated in pure noble gases.

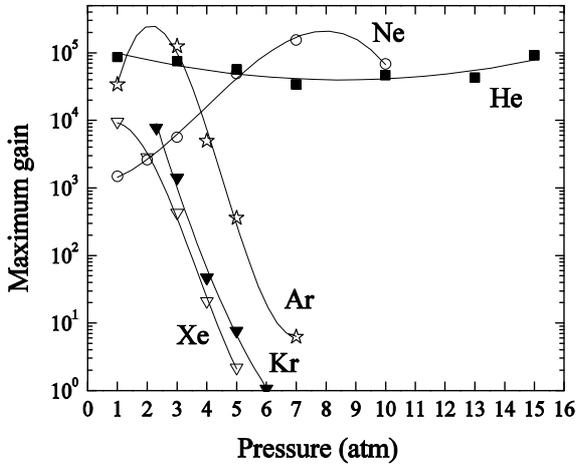

Fig.2 Maximum gain of a triple-GEM detector as a function of pressure in He, Ne, Ar, Kr and Xe [9,10].

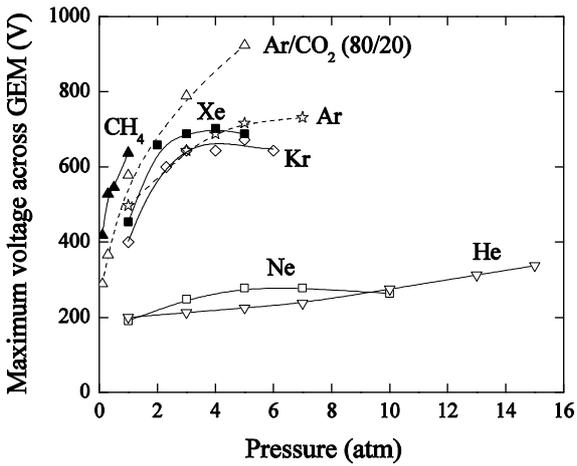

Fig.3 Maximum operation voltage that can be applied across each GEM in a triple-GEM detector as a function of pressure, in He, Ne, Ar, Kr and Xe [9,10]. The data obtained in Ar/CO$_2$ and CH$_4$ in a single-GEM detector [25] are presented for comparison.

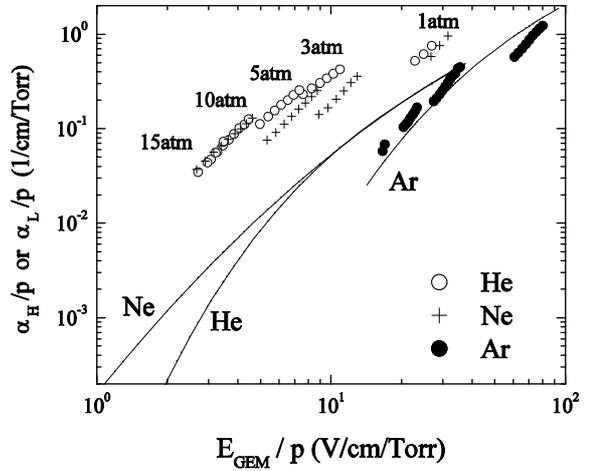

Fig.4 Comparison of ionization coefficients at high ($\alpha_H$, points) and low ($\alpha_L$, curves) pressures, in He, Ne and Ar. The $\alpha_H$ values were obtained using the triple-GEM detector gain at appropriate electric field inside the GEM hole, $E_{GEM}$. The $\alpha_L$ values were compiled from the literature.

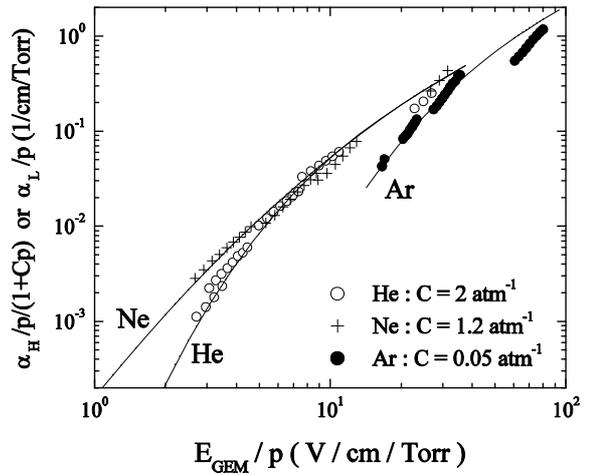

Fig.5 Fit of the quantity $\alpha_H/p/(1+Cp)$ obtained from the experimental data at high pressures (points) to the reduced ionization coefficients at low pressures, $\alpha_L/p$, taken from the literature (curves). Parameter $C$ describes the contribution of the associative ionization in He, Ne and Ar.

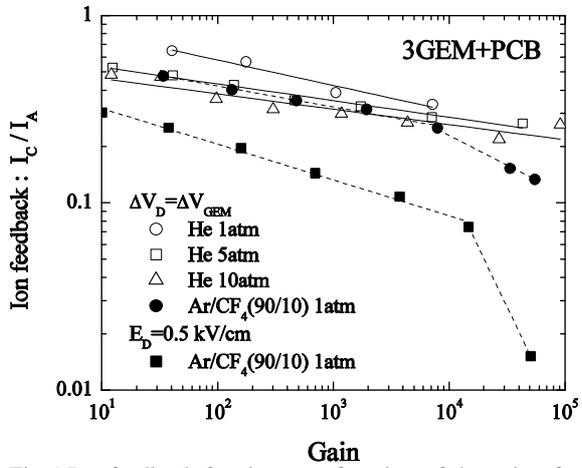

Fig.6 Ion feedback fraction as a function of the gain of a triple-GEM detector, in He at 1, 5 and 10 atm and in Ar/CF$_4$ at 1 atm [7]. Two data sets are shown: with a drift field proportional to electric field inside the GEM hole ($\Delta V_D = \Delta V_{GEM}$) and with a constant drift field ($E_D$=0.5 kV/cm).